# Photoacoustic-guided reflection-mode wavefront shaping through dynamic biological tissue

**Xuan Liu,[1] Sacha Noimark,[2,3] Wenfeng Xia,[1] Sebastien Ourselin,[1] and Tianrui Zhao[1,*]**
[1]School of Biomedical Engineering and Imaging Sciences, King's College London, 4th Floor, Lambeth Wing, St Thomas' Hospital, London SE1 7EH, United Kingdom
[2]Department of Medical Physics and Biomedical Engineering, University College London, Malet Place Engineering Building, London, WC1E 6BT, United Kingdom
[3]Department of Chemistry, University College London, 20 Gordon Street, London, WC1H 0AJ, United Kingdom
[*]Correspondence: tianrui.zhao@kcl.ac.uk

## GRAPHICAL ABSTRACT

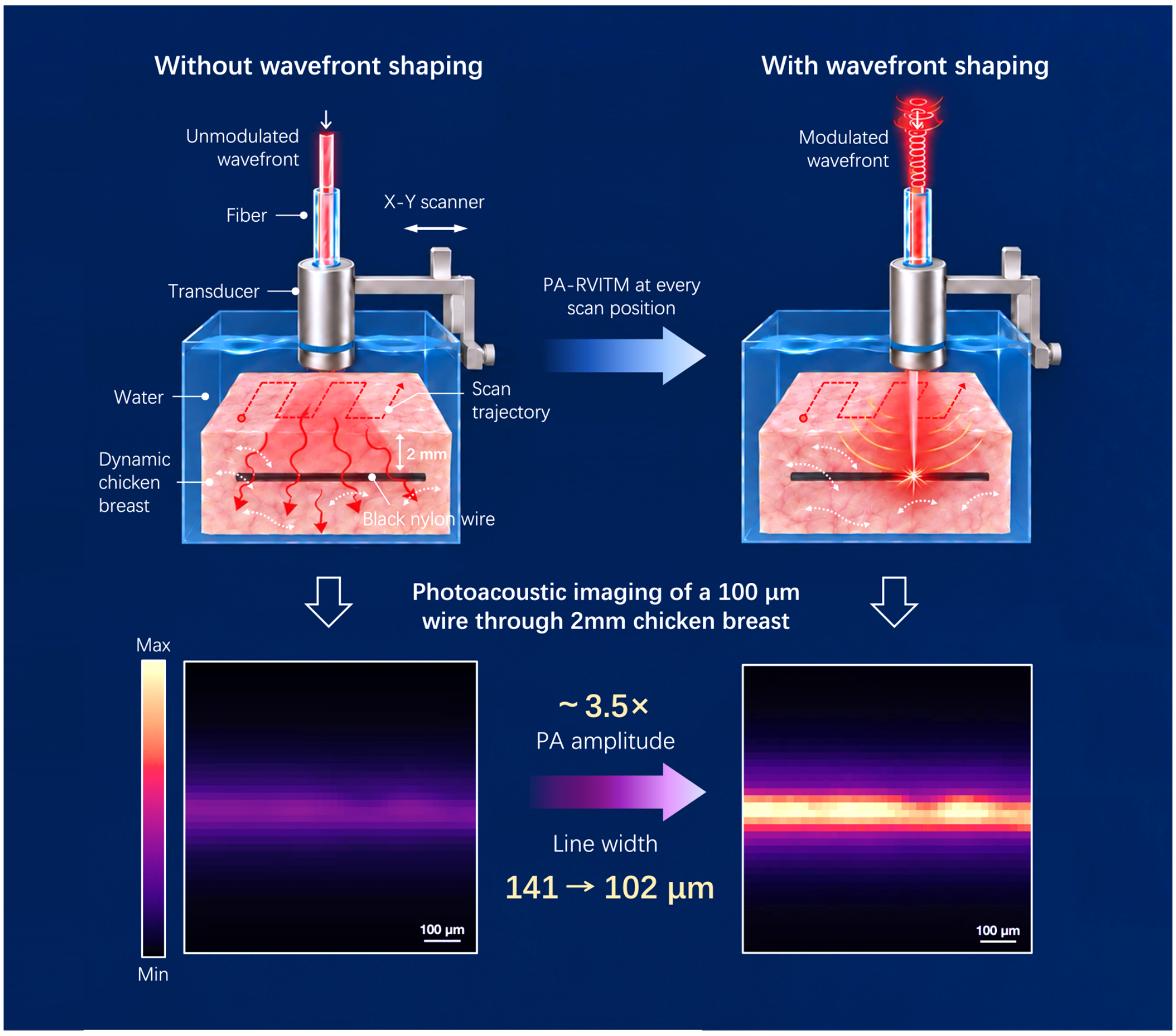

## PUBLIC SUMMARY

- Photoacoustic (PA)-guided wavefront shaping focuses light deep in tissue using sound as feedback.
- Existing systems need optical and acoustic access from opposite sides, limiting them to thin samples.
- We developed a reflection-mode PA-guided wavefront shaping method with fiber delivery and sub-second focusing.
- It boosted the PA signal ~14-fold through a static diffuser and ~3.5-fold through 2 mm

of chicken breast.

- We achieved PA microscopy of a 100-μm wire through 2 mm of *ex vivo* tissue with enhanced signal and contrast.

## ABSTRACT

Strong optical scattering limits the depth at which light can be focused and used for imaging in biological tissue. Photoacoustic (PA)-guided wavefront shaping enables non-invasive light focusing in scattering media, but most existing implementations require optical and acoustic access from opposite sides of the sample and a rigidly aligned free-space optical path, which have confined these demonstrations to thin or static samples. Here, we combine a real-valued intensity transmission matrix (RVITM) algorithm with a photoacoustic guidestar (PA-RVITM) in a reflection-mode configuration. Modulated light is delivered through a flexible multimode fiber passing through the central aperture of a focused ultrasound transducer, enabling optical illumination and acoustic detection from the same side of the sample without a free-space combiner. Fiber integration preserved the wavefront shaping performance, and the reflection-mode PA-RVITM system achieved sub-second wavefront optimization with a ~14-fold PA signal enhancement through a diffuser. In *ex vivo* chicken breast tissue, the enhancement decreased with tissue thickness, reaching ~3.5 through a 2-mm chicken breast layer, with a speckle decorrelation time of ~450 ms. By performing an independent RVITM optimization at each scan position, our system enabled photoacoustic microscopy of a black nylon wire through the same layer, reducing the imaged line width from ~141 μm to ~102 μm, close to the nominal diameter of ~100 μm. This work advances PA-guided wavefront shaping toward a compact, same-side implementation for future biomedical therapeutic and imaging applications.

## KEYWORDS

Wavefront shaping, transmission matrix, tissue dynamics, biological tissue, photoacoustic microscopy, reflection mode

## INTRODUCTION

Manipulating light deep inside biological tissue is important for a broad range of biomedical applications, ranging from light-driven interventions such as optogenetic stimulation, microsurgery, and photodynamic therapy, to diagnostic imaging including cancer detection, brain imaging, and molecular imaging.[1,2] However, the depth at which light can be focused is fundamentally limited by the microscopic inhomogeneity of tissue: multiple scattering randomizes photon trajectories and distorts the incident wavefront, making it difficult to focus light beyond the optical diffusion limit (~1 mm), precluding the formation of a clear image of deep tissue.[3,4]

Wavefront shaping has been extensively explored to overcome this limitation.[5] Although multiple scattering scrambles the incident light into a random speckle field, light propagation through a static scattering medium is a deterministic and linear process. By pre-shaping the incident wavefront using a spatial light modulator (SLM), the scattering-induced distortions can be compensated, thereby enabling deep optical focusing through coherent superposition of scattered light at a target location.[6–8] Wavefront shaping typically relies on feedback from transmitted light measured by optical detectors positioned behind the scattering medium.[9–11] However, in biological tissues, such access is challenging because neither the tissue interior nor its far side is generally accessible to an optical detector. To provide internal feedback without direct optical access, various 'guidestars' have been developed, including fluorescent,[12–15] magnetic particle-based,[16] ultrasonic,[17–19] and photoacoustic (PA) guidestars.[20–28]

PA guidestars have attracted considerable interest because they eliminate the need for exogenous guide particles while enabling feedback from greater depths through low-attenuation acoustic detection.[29] Moreover, the enhanced PA signal resulting from optical focusing can directly improve PA imaging performance. In a typical PA-guided wavefront shaping implementation, a focused ultrasound transducer detects PA signals generated at its acoustic focus, where the PA amplitude is proportional to the local optical fluence. Optical focusing is achieved by maximizing the detected PA signal through iterative incident wavefront modulation. In 2011, Kong et al.[28] first used a PA guidestar to focus light through a diffuser, and in 2013, Chaigne et al.[27] introduced a three-phase-shift algorithm that achieved selective focusing within scattering media. Over the following decade, the resolution and signal-to-noise ratio (SNR) of PA-guided wavefront shaping were progressively improved through nonlinear PA responses,[22] non-uniform transducer sensitivity,[21] lock-in detection,[23] and wavelet denoising.[25]

Despite these advances, a major barrier to practical imaging is that most PA-guided wavefront-shaping studies employ a transmission geometry, with optical illumination and acoustic detection positioned on opposite sides of the scattering medium (Figure 1A).[21–27,30,31] Although transmission-mode configurations can be used for imaging and are well suited to proof-of-concept studies, they are often impractical for clinical applications, where access to both sides of the tissue is limited and imaging systems are typically operated from a single accessible surface. This is particularly restrictive for deep organs and lesions, for which positioning illumination and detection components on opposing sides of the body may be infeasible. In addition, light must propagate through a large tissue thickness before reaching the acoustic focus, resulting in substantial attenuation and weak PA feedback for wavefront optimization. A reflection-mode configuration, with optical illumination and acoustic detection on the same side of the tissue, is therefore far more compatible with practical clinical PA imaging (Figure 1B). In photoacoustic microscopy (PAM), same-side optical illumination and acoustic detection are already well established, typically implemented using a prism–silicone-oil combiner to align the optical and acoustic axes coaxially (Figure 1C).[32–36] However, integrating such a coaxial PAM architecture with a wavefront-shaping module results in a bulky and mechanically complex confocal assembly. Because conventional PAM relies on spatial scanning of the optical–acoustic focus to form an image, translating or raster-scanning this enlarged assembly is technically challenging and limits its suitability for practical implementation.

In this work, we developed a reflection-mode geometry (Figure 1D) based on an optical fiber and a through-hole ultrasound transducer and demonstrated light focusing through scattering media using a PA-guided real-valued intensity transmission matrix (PA-RVITM). We first validated the reflection-mode PA-RVITM system by focusing light through static scattering media, and then extended the approach to dynamic media using *ex vivo* chicken tissue. Finally, taking advantage of the compact probe geometry and its ease of spatial scanning, we demonstrated enhanced PA imaging with wavefront-shaped illumination.

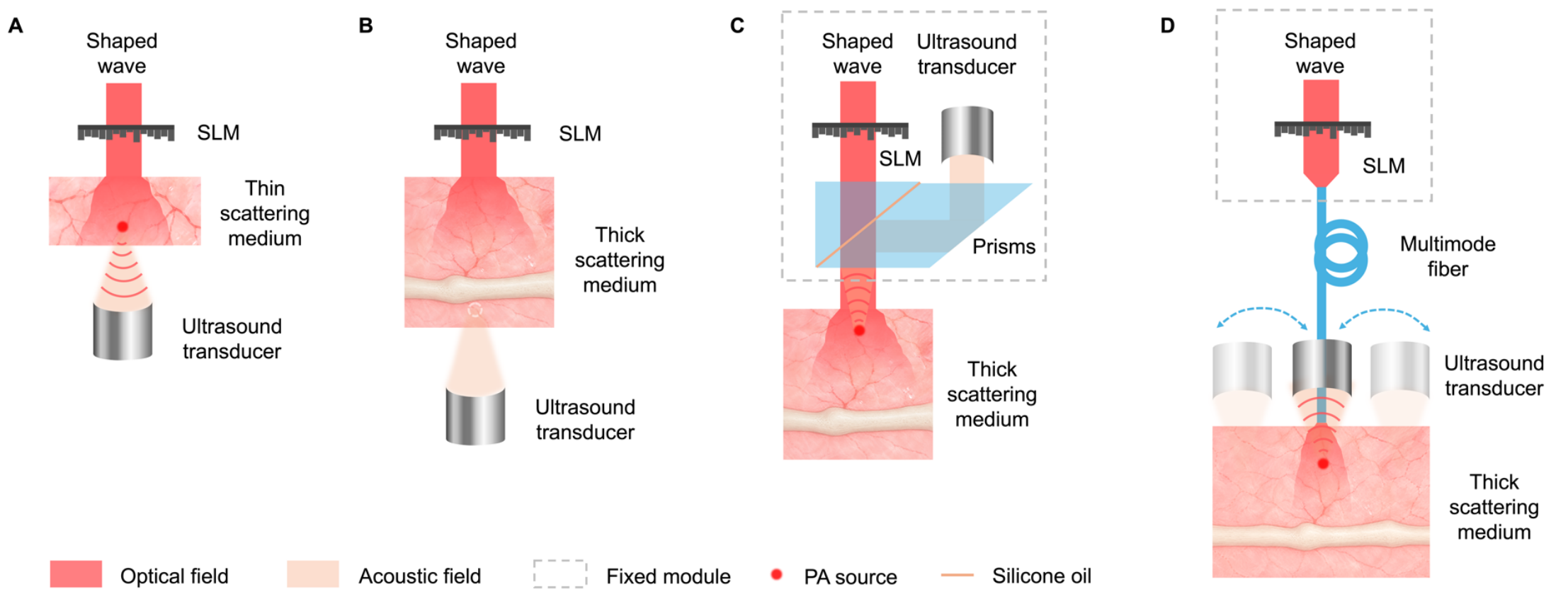

**Figure 1. Configurations for photoacoustic (PA)-guided wavefront shaping.** (A) Conventional opposite-side geometry with a thin scattering medium: a spatial light modulator (SLM) shapes the incident light and a focused ultrasound transducer on the far side detects the PA signal generated at the focus. (B) The same geometry with a thick scattering medium, where strongly attenuating structures such as bone absorb and scatter the light before it reaches the transducer side, hindering effective PA feedback. (C) Prism-based coaxial confocal probe: a silicone-oil layer between two prisms combines the optical and acoustic beams for same-side operation, but the optics and transducer form a single rigid module (dashed box). (D) Same-side, fiber-integrated configuration of PA-RVITM used in this work: shaped light is delivered through a multimode fiber and the central through-hole of the transducer. Only the SLM remains fixed (dashed box), while the flexible probe can be repositioned without realigning the benchtop modulation optics (arrows).

## MATERIALS AND METHODS

### Ultrasound transducer design

A spherically focused ultrasound transducer was custom-made (Precision Acoustics, UK) to support coaxial optical illumination and acoustic detection (Figure 2A). The transducer used a polyvinylidene fluoride (PVDF) active element, selected for its mechanical flexibility, ease of machining for the central aperture, and compatibility with the curvature required for acoustic focusing.[37] The transducer had a diameter of 22 mm and a 0.5-mm-diameter through-hole drilled along the optical axis. The acoustic characterization data shown in Figures 2B–E were measured by Precision Acoustics and reproduced here. The measured time-domain response and corresponding frequency spectrum confirmed a center frequency of ~13 MHz (Figures 2B–C). The −6 dB focal dimensions were approximately 80 × 90 µm laterally and 380 µm axially (Figures 2D–E), with the focus located approximately 5 mm from the transducer surface. The aperture size of 0.5 mm was selected to accommodate a 300-µm multimode fiber with sufficient mechanical clearance, while minimizing the reduction in active aperture area. This configuration enabled direct optical delivery through the center of the transducer and PA signal detection from the same side of the sample.

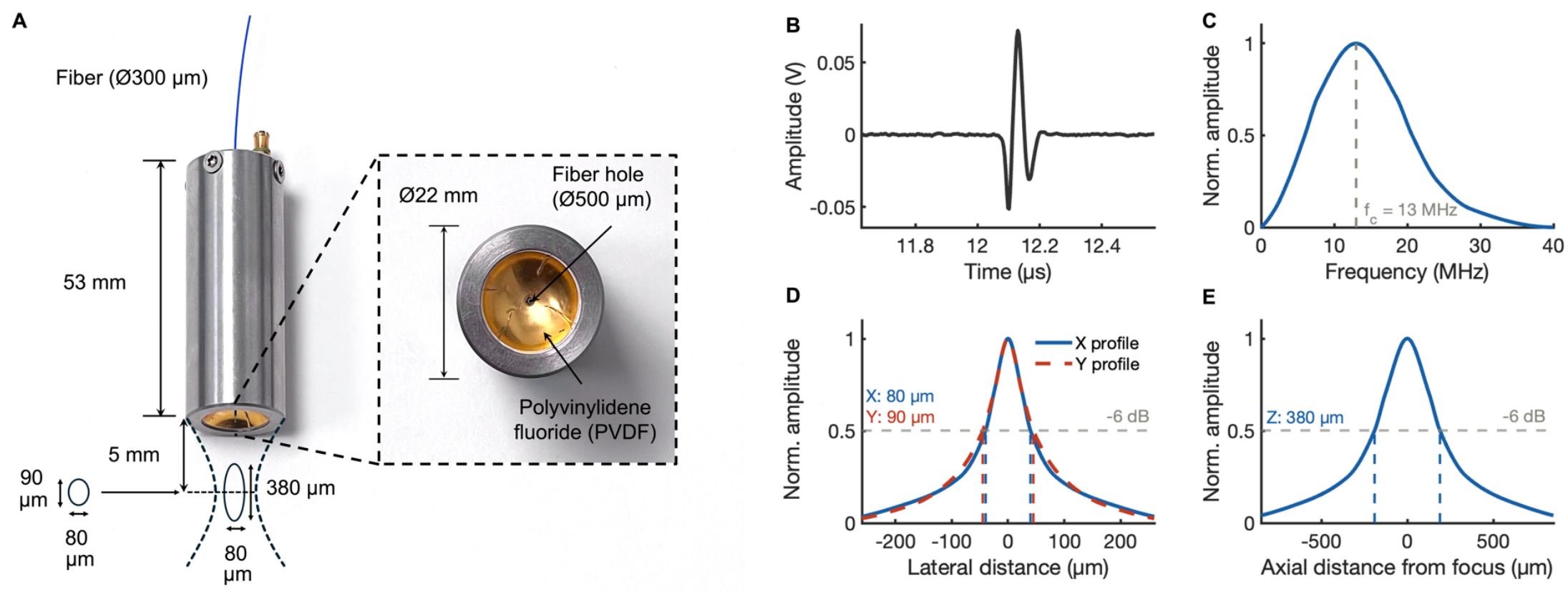

**Figure 2. Structure and acoustic characterization of the customized through-hole transducer.** (A) Photograph and dimensions of the customized transducer, showing the central fiber through-hole and PVDF active element. (B) Measured time-domain response. (C) Corresponding frequency spectrum. (D) Normalized lateral profiles along the x and y directions. (E) Normalized axial profile.

### Experimental setup

The wavefront modulation module is illustrated in Figure 3. A pulsed laser (Onda 1064, BrightSolutions, Cura Carpignano, PV, Italy) operating at 1064 nm with a pulse width of 5 ns was used for PA signal excitation, and the beam was spatially modulated by a digital micromirror device (DMD) (DLP650LNIR, 1280 × 800 pixels, Texas Instruments, USA) operating at 20 kHz. A total of 4096 input modes were used, with each mode formed by grouping 16 micromirrors into one macro-pixel on the DMD. The modulated light was coupled into an optical fiber (Ø300 µm, 0.39 NA, 1 m, M56L01, Thorlabs, New Jersey, USA) via a convex lens (AC254-030-A-ML, Thorlabs, New Jersey, USA) and an objective lens (20×, 0.4 NA, RMS20X, Thorlabs, New Jersey, USA), and delivered through the central aperture of the transducer to the scattering medium. The detected PA signal was amplified and digitized using a data acquisition card (M4i.4420, Spectrum Instrumentation, Germany) and processed on a personal computer (Intel i7 processor, 3.2 GHz) to provide feedback for the RVITM algorithm. Laser pulses and DMD pattern display were synchronized by an arbitrary waveform generator (33600A, Keysight Technologies, USA). PA signals were processed in MATLAB with a 15 MHz low-pass filter to suppress high-frequency noise.

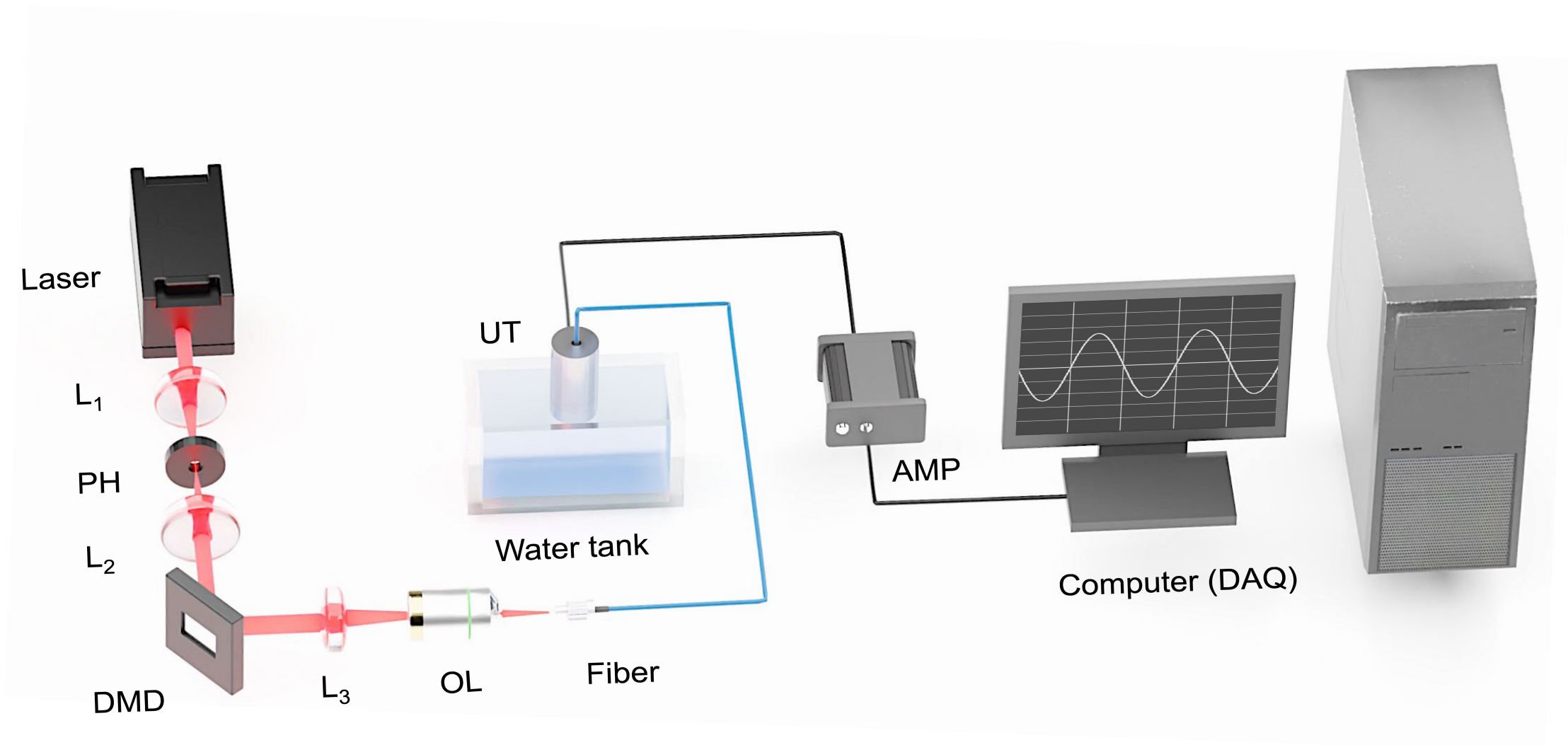

**Figure 3. Experimental setup of the** photoacoustic-guided reflection-mode RVITM wavefront shaping system**.** $L_{1–3}$, convex lenses; PH, pinhole; DMD, digital micromirror device; OL, objective lens; UT, ultrasound transducer; AMP, amplifier; DAQ, data acquisition system.

### Photoacoustic-guided RVITM

The RVITM algorithm has been detailed in our prior studies.[26,38] Briefly, a Hadamard matrix H ∈ {−1, +1} was used to generate the complementary binary matrices $H_1$ = (H + 1)/2 and $H_2$ = (−H + 1)/2; each of the 2*N* columns of [$H_1$, $H_2$] was reshaped into a binary DMD pattern and projected in sequence while the corresponding PA amplitude was recorded. Since PA amplitude is proportional to the local optical fluence at the acoustic focus ( $P^k = \alpha I^k$, where α incorporates detector sensitivity, acoustic attenuation, and the Grüneisen parameter; $I^k$ is the local light intensity at the target produced by the *k*th DMD pattern), PA signal amplitude can be directly used as the output intensity. Then, RVITM builds a pseudo-linear link between input and output modes in intensity, with each transmission element linking the *n*th input mode to the PA amplitude expressed as:

$$\mathrm{RVITM}_{\mathrm{PA}}^{\mathrm{n}} = \alpha N A_R A_n \cos(\theta_n - \varphi_R)\,, \tag{1}$$

where $A_n$ and $\theta_n$ are the amplitude and phase of the transmission coefficient from the *n*th input mode to the target, $A_R$ and $\varphi_R$ are those of the reference field at that location, and *N* is the number of input modes. Thus, the sign of each coefficient is set by $\cos(\theta_n - \varphi_R)$, indicating whether the corresponding input mode contributes constructively or destructively relative to the reference field at the target. To maximize $P^k$, the optimal DMD pattern was obtained by switching on all input modes with positive coefficients.

### Effect of fiber delivery on focusing performance

To assess whether multimode-fiber delivery affected the wavefront shaping performance, two optical configurations were compared: a “diffuser-only” configuration and a “fiber + diffuser” configuration (Figure 4A). In the “diffuser-only” configuration, the DMD-modulated field illuminated the diffuser (N-BK7 Ground Glass Diffuser, 220 Grit, Thorlabs, New Jersey, USA) directly. In the “fiber + diffuser” configuration, the modulated field was first transmitted through the multimode fiber and then through the same diffuser.

Feedback for the RVITM optimization was provided by a fiber-coupled photodetector (APD410A/M, Thorlabs, New Jersey, USA) placed behind the diffuser.[39] The collection fiber of this photodetector had a core diameter of 100 μm, chosen to be comparable to the lateral dimensions of the acoustic focal region, so that the optical-feedback area approximated the spatial region sampled by the PA guidestar in the subsequent experiments. The laser energy and total number of input modes ($N$ = 4096) were kept constant between the two configurations. Measurements were repeated nine times for each configuration. After optimization, the collection fiber was temporarily replaced by a camera (acA1300-200um, Basler, Germany) positioned at the same target plane to record the speckle and focused intensity distributions. The enhancement factor (η) and full width at half maximum (FWHM) of the focus were quantified from the recorded speckle and focused images. The enhancement η was defined as the ratio of the focal intensity obtained with the optimal pattern to the mean intensity obtained with random patterns, and the FWHM was obtained by Gaussian fitting of the lateral focal profile.

**Photoacoustic-guided light focusing**

To evaluate the focusing performance of the reflection-mode PA-RVITM system, a 1-cm-wide black tape absorber (3M, USA) was sandwiched between two static diffusers and positioned at the acoustic focus of the transducer (Figure 5A). The fiber–transducer assembly was immersed in a water tank for coupling. Light delivered through the fiber illuminated the absorber through the first diffuser, generating PA signals that were detected by the transducer and amplified by a preamplifier (SPA.1411, Spectrum Instrumentation, Germany). The peak-to-peak PA amplitude was used as the optimization feedback. Signals acquired under random and optimized conditions were normalized for comparison.

**Photoacoustic imaging through *ex vivo* tissue**

To evaluate the reflection-mode PA-RVITM system in a dynamic biological medium, *ex vivo* chicken breast tissue was used as a representative scattering medium. A black tape absorber was embedded between two chicken breast slices of equal thickness to mimic a localized absorber in biological tissue (Figure 6A). Throughout this study, the reported tissue thickness refers to the thickness of the overlying chicken breast layer between the probe and the absorber. For each condition, the thickness of both tissue slices was varied together from 1 to 5 mm. Under identical experimental conditions, PA signals were then measured with and without wavefront shaping for each thickness, and the corresponding enhancement factors were calculated. Tissue decorrelation time ($\tau_c$) was estimated from the temporal autocorrelation of sequential speckle images and defined as the time at which the normalized autocorrelation decayed to $1/e^2$.[40] Optical energy delivery was measured using the fiber-coupled photodetector positioned at the absorber plane. The incident pulse energy was measured under the same geometry without tissue, and the delivery efficiency was calculated as the transmitted energy at the target plane divided by the corresponding incident energy. The input energy was maintained constant across tissue thicknesses. For each thickness, all three quantities were acquired at nine spatially separated positions.

PA imaging was performed in the same configuration with a 2-mm overlying tissue layer, the thickness at which focusing had been characterized above. A black nylon wire (Suturing Doctor, China) was used as the absorbing target and embedded between two 2-mm-thick chicken breast layers (Figure 7A). The wire diameter was independently measured by optical microscopy as approximately 100 μm. The fiber–transducer probe was raster-scanned across the nylon wire using a motorized XY translation stage (M30XY/M, Thorlabs, New Jersey, USA), with a step size of 20 μm over an 800 × 800 μm field. At every scan position, an independent RVITM optimization (4096 input modes) was performed to obtain the locally optimal wavefront, and the peak-to-peak PA amplitude at that position was recorded to generate the optimized PAM image. For comparison, an unoptimized PAM image of the same wire at the same positions was acquired under identical laser energy and scanning parameters using random wavefronts, with the incident wavefront as the only differing variable between the two conditions. Profiles across the wire were extracted perpendicular to the wire axis after background subtraction. The FWHM was calculated for each scan column using direct half-maximum interpolation, and the resulting values were summarized using the median and interquartile range across all columns.

## RESULTS

### Fiber validation

The focusing performance was evaluated under two optical configurations: "diffuser-only" and "fiber + diffuser" (Figure 4A). Boxplots of enhancement factors measured from nine repeated experiments for each configuration are shown in Figure 4B. Representative speckle patterns, focused images, and the lateral profiles are presented in Figures 4C–D. Similar mean optical-feedback enhancement factors (122 vs. 120) and similar focal FWHM values (99.2 μm vs. 98.6 μm) were obtained from the two configurations, with comparable distributions across repeats. Both FWHM values were comparable to the 100-μm core diameter of the detection fiber. These results indicate that introducing a multimode fiber has little impact on wavefront shaping performance.

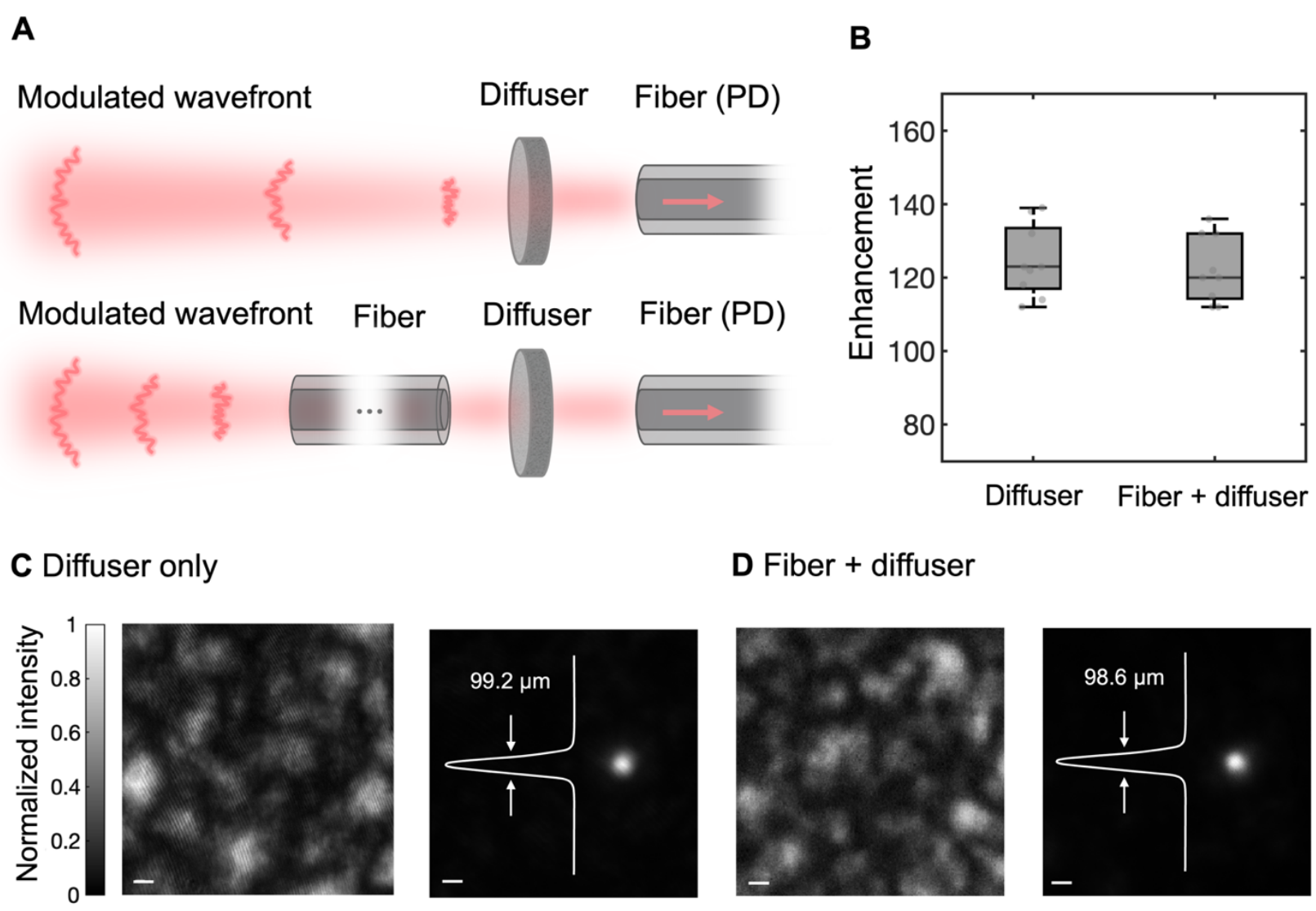

**Figure 4. Comparison of wavefront optimization with and without multimode fiber delivery, measured with optical feedback.** (A) Experimental configurations. A nominal 100-μm-core collection fiber coupled to a photodetector (PD) provided optical-feedback measurements. (B) Enhancement factors obtained from nine repeated measurements in each configuration. (C) Representative speckle pattern, optimized focus, and lateral intensity profile for the diffuser-only configuration. (D) Corresponding results for the fiber + diffuser configuration. Profile widths reflect the collection aperture rather than the true focal spot size. Scale bars are 100 μm.

## Photoacoustic-guided light focusing

The static performance of the reflection-mode PA-RVITM system through a diffuser is shown in Figure 5. A black tape absorber was placed between two diffusers and positioned at the acoustic focus of the transducer (Figure 5A). After wavefront optimization, the PA signal amplitude increased approximately 14-fold compared with that obtained using random wavefront patterns (Figure 5B), demonstrating successful closed-loop focusing using the developed system with fiber-delivered modulated light and a PA guidestar. The total optimization cycle was completed in 467 ms, with the time required for each step in the workflow shown in Figure 5C. Before the characterization step, 8192 (2*N*) input DMD patterns were processed and uploaded into the DMD's memory. The most time-consuming process was displaying these DMD patterns at a rate of 20 kHz (410 ms) while simultaneously acquiring the PA signal. An additional 19 ms was required to transfer the PA signal from the DAQ card to the PC memory. The optimal DMD pattern was then computed in 26 ms with a custom MATLAB implementation of the RVITM algorithm. The final step involved uploading and displaying the optimal pattern on the DMD for light focusing, which took 12 ms. These results demonstrate the functional integration of fiber delivery, PA feedback, and RVITM-based wavefront optimization within the same-side system.

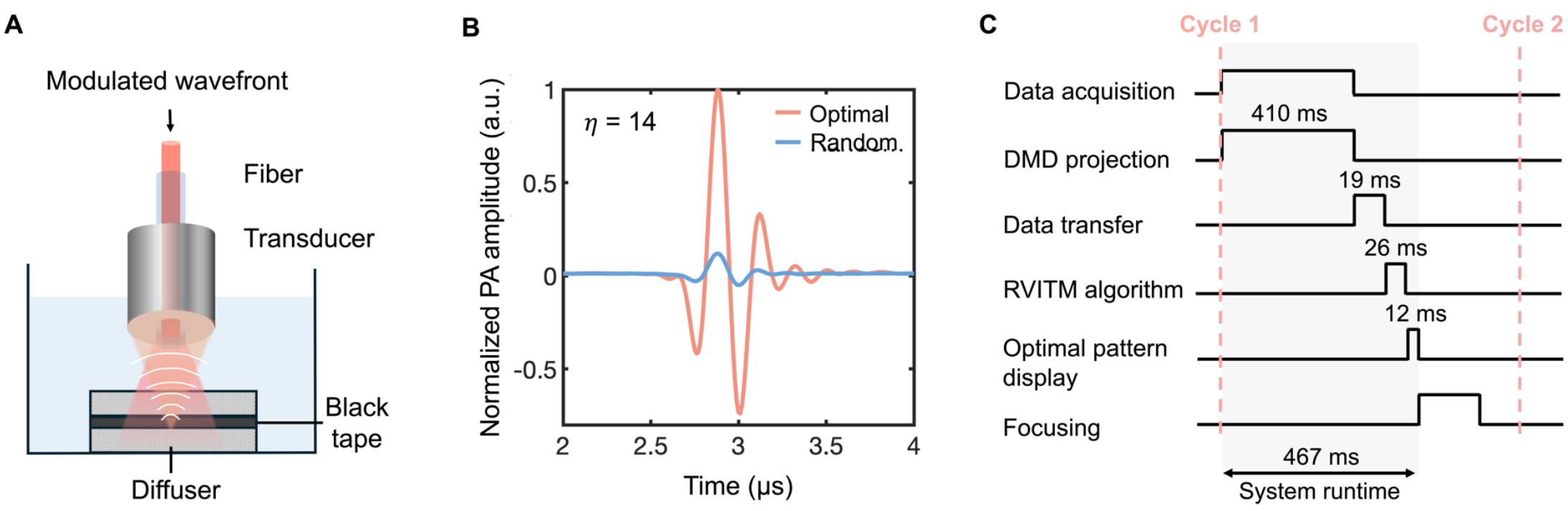

**Figure 5. Static performance and timing characterization of the photoacoustic-guided reflection-mode RVITM wavefront shaping system.** (A) Schematic of the characterization setup, with a black-tape absorber placed between two diffusers. (B) Representative normalized PA waveforms obtained with the optimal and random modulation patterns, showing a PA enhancement factor of 14. (C) Timing sequence and runtime breakdown of one PA-guided RVITM optimization cycle using 4096 DMD input modes, with a total optimization cycle of 467 ms.

## Photoacoustic imaging through *ex vivo* tissue

Following the static-diffuser experiments, the reflection-mode PA-RVITM system was further evaluated in *ex vivo* chicken breast tissue, a dynamic and optically thick medium (Figure 6A). Representative waveforms showed enhancement factors of 4.6, 3.1, and 1.0 at tissue thicknesses of 1, 3, and 5 mm, respectively (Figure 6B). Across nine repeated measurements, the mean enhancement decreased progressively with increasing overlying tissue thickness, from ~4.4 at 1 mm to ~3.5 at 2 mm and ~1.0 at 5 mm (Figure 6C). Two additional thickness-dependent quantities were measured. The speckle decorrelation time decreased from approximately 530 ms at 1 mm to 450 ms at 2 mm and 300 ms at 5 mm (Figure 6D). The mean decorrelation time exceeded the 467 ms optimization cycle at 1 mm, was comparable to it at 2 mm, and fell below it at larger thicknesses. The optical energy delivery efficiency decreased from approximately 1.5% at 1 mm to 0.2% at 5 mm (Figure 6E).

Based on the point-focusing results, we examined whether the developed system could image an absorbing target through *ex vivo* tissue. A tissue thickness of 2 mm was selected because it was the maximum thickness at which wavefront shaping produced a clearly enhanced and continuous wire image. A black nylon wire (nominal diameter, ~100 μm) was embedded between two 2-mm-thick chicken breast layers and imaged with and without wavefront shaping (Figure 7). Under random, unoptimized wavefronts, the target appeared only weakly against the background (Figure 7C), whereas with an independent RVITM optimization at every scan position the wire appeared as a continuous line across the full 800 × 800 μm field (Figure 7D). Wavefront shaping increased the PA amplitude along the wire, thereby improving contrast against the background. The measured line width (FWHM) decreased from a median of ~141 μm without wavefront shaping to ~102 μm with wavefront shaping, close to the wire diameter measured by optical microscopy (Figure 7B). The per-column FWHM values obtained with wavefront shaping were also less dispersed than those obtained without wavefront shaping (Figure 7E). The broader width measured under random illumination is consistent with the acoustic response of the assembled probe, for which the nominal lateral focal dimensions were characterized on the transducer alone, without the fiber inserted in the central aperture. Wavefront optimization may then narrow the measured response by concentrating the incident light within a smaller portion of the region sampled by the transducer, thereby confining the effective optical excitation.

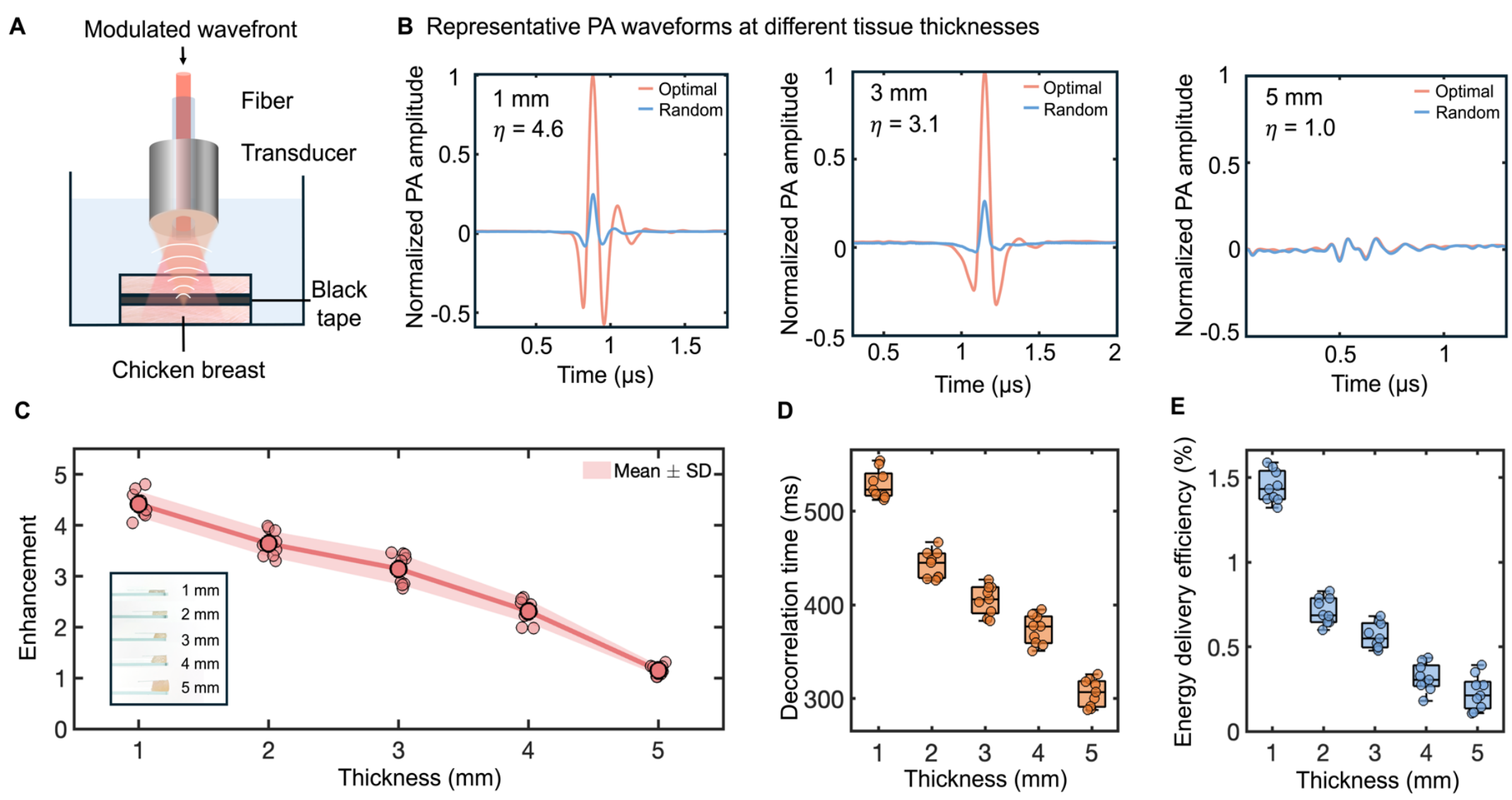

**Figure 6. Depth dependence of the reflection-mode PA-RVITM system performance in *ex vivo* chicken breast tissue.** (A) Schematic of the measurement configuration, with a black-tape absorber placed between two tissue layers of equal thickness. (B) Representative normalized PA waveforms obtained with the optimal and random modulation patterns at tissue thicknesses of 1, 3, and 5 mm. (C) PA signal enhancement factor (η) measured at different tissue thicknesses. (D) Speckle decorrelation time ($\tau_c$) measured at different tissue thicknesses. (E) Optical energy delivery efficiency at different tissue thicknesses.

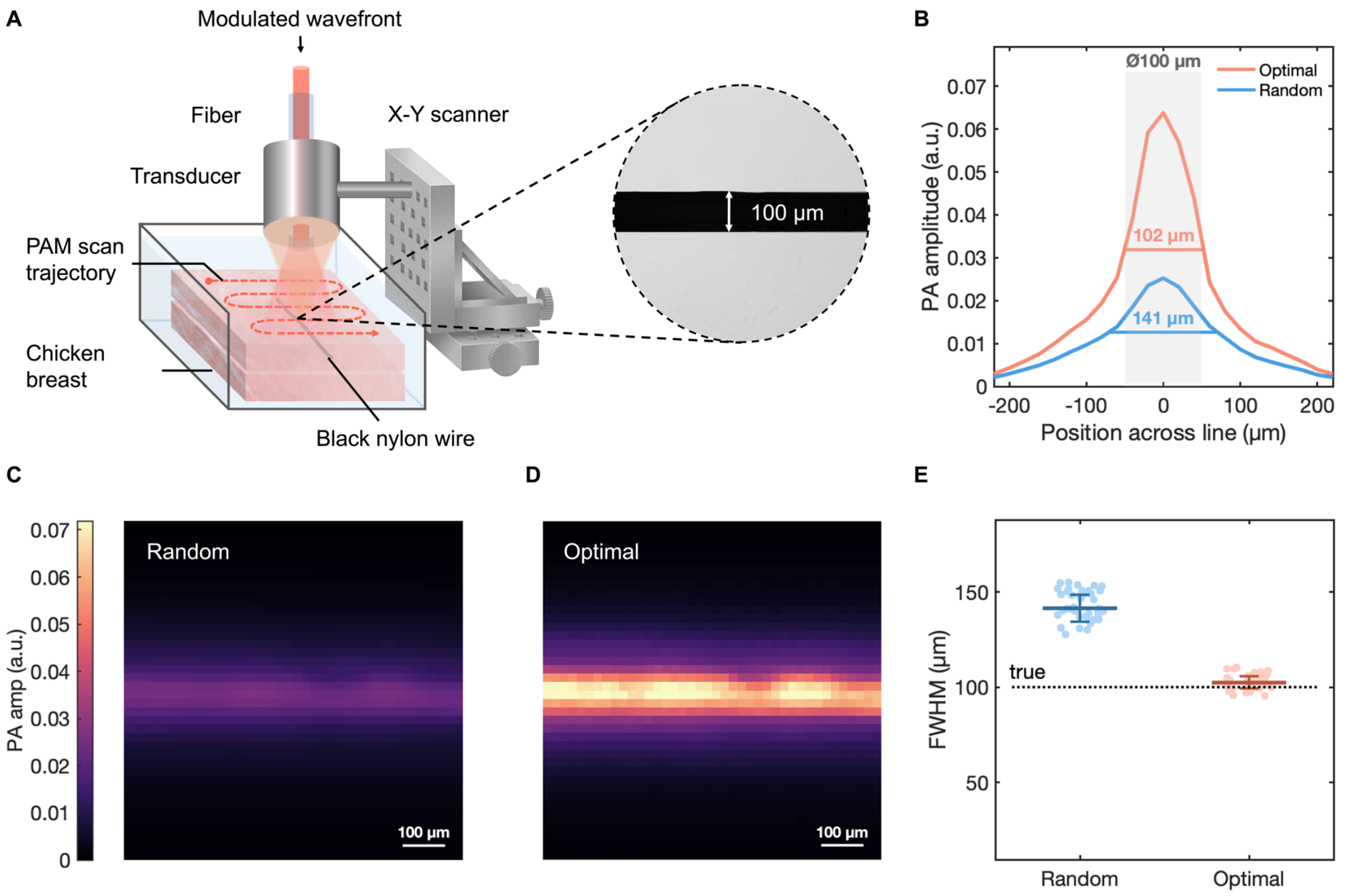

**Figure 7. Photoacoustic microscopy of a black nylon wire through a 2-mm *ex vivo* tissue layer with per-position PA-RVITM optimization.** (A) Schematic of raster scanning of the fiber–transducer probe across the wire, with an inset micrograph of the wire (nominal ~100 μm). An independent RVITM optimization was performed at every scan position. (B) Representative background-subtracted PA amplitude profiles across the wire obtained with optimized and random wavefronts. (C) PAM image acquired with unoptimized random wavefronts. (D) PAM image acquired with per-position RVITM optimization. (E) Distribution of the column-wise wire FWHM. Markers indicate the median and horizontal bars the interquartile range across all columns. The dashed line marks the nominal wire diameter.

## DISCUSSION

This study demonstrated a reflection-mode PA-RVITM system in which optical illumination and acoustic detection are performed from the same side of the scattering medium. This geometry is more compatible with practical PA imaging than the transmission-mode configurations commonly used in previous studies, particularly for biological tissues where access to the opposite surface may be limited. By integrating an optical fiber with a through-hole ultrasound transducer, the system also provides a compact probe architecture that can be readily translated for image acquisition. Fiber integration decouples the sensing head from the DMD and relay optics, allowing the PA-RVITM probe to be repositioned for scanning while the modulation optics remain stationary. Comparable enhancement factors and focal dimensions obtained with and without fiber delivery indicate that wavefront control is preserved after fiber integration. As long as the fiber remains stable during sub-second characterization at each scan position, its mode coupling forms part of the overall transmission response characterized by RVITM.

Temporal changes in the scattering medium continuously alter the optical transmission characteristics, causing an optimized wavefront to lose validity once the medium decorrelates. PA-guided wavefront shaping in the multiple-scattering regime has so far been demonstrated in static media, using methods that are too slow for a decorrelating sample: feedback-based iterative methods require numerous sequential measurements to converge,[8,28,41] and conventional PA-guided transmission-matrix characterization requires at least 3*N* phase-shifting measurements for *N* input modes.[27] Optical phase conjugation (OPC) is fast enough and can focus light through thick tissue,[42–44] but it cannot be combined with a PA guidestar, as phase conjugation requires complex-field measurements while a transducer reports amplitudes only. With an ultrasonic guidestar, OPC uses the ultrasound-tagged light as feedback, which must be collected optically.[18] Even reflection-mode implementations require the acoustic axis to remain separate from the optical path, restricting the method to samples accessible from more than one direction.[45,46] RVITM instead employs the PA amplitude directly as the intensity feedback and requires only 2*N* intensity measurements. Combined with high-speed DMD modulation, one PA-RVITM optimization cycle was completed in 467 ms, allowing the reconstructed RVITM to remain sufficiently valid even through the dynamic 2-mm-thick tissue.

On a static diffuser, PA-RVITM achieved an approximately 14-fold PA signal enhancement, which is comparable to values reported in previous studies using linear PA feedback.[21,23,25,27,47–52] When the static diffuser was replaced by 2-mm-thick chicken breast tissue, the enhancement decreased to approximately 3.5-fold and vanished at 5 mm. On the one hand, stronger optical attenuation yields weaker PA signals, which reduce the accuracy of the RVITM reconstruction and thereby limit the achievable enhancement. This interpretation is supported by the substantially larger enhancement obtained under high-SNR optical-feedback conditions and previous reports showing the dependence of PA-guided wavefront optimization on feedback quality.[24] On the other hand, increasing tissue thickness results in more extensive multiple scattering, making the transmitted speckle field increasingly sensitive to microscopic scatterer motion and thereby accelerating temporal decorrelation.[53,54] Consequently, the effective transmission matrix becomes increasingly outdated during optimization, leading to a further reduction in the achievable PA enhancement.

Furthermore, in this proof-of-concept evaluation under controlled biological conditions, *ex vivo* chicken breast does not reproduce the faster decorrelation and physiological motion observed in living tissue, while the strongly absorbing nylon wire provided more favorable PA feedback than many endogenous absorbers. Future studies could evaluate the SNR of endogenous PA feedback and system performance under physiological dynamics. Improving the achievable enhancement factor will be important for *in vivo* imaging applications. PA feedback SNR could be improved by using transducers based on higher-sensitivity piezoelectric materials, together with improved pulse-to-pulse laser stability and increased incident pulse energy within applicable safety limits.[23,25,55] Faster modulators, such as micro-electromechanical system (MEMS)-based phase modulators and one-dimensional grating light valves,[56,57] together with field-programmable gate array (FPGA)-based signal processing,[11,14] could shorten the optimization cycle. Alternatively, reducing the mode number would accelerate optimization, but at the cost of fewer optical degrees of freedom and a lower attainable enhancement.[58,59]

## CONCLUSION

In summary, we demonstrated a reflection-mode PA-RVITM system, in which PA-guided wavefront shaping is performed with optical illumination and acoustic detection on the same side of the sample, using a flexible fiber-integrated probe rather than a rigidly aligned free-space path. This method removes the requirement for distal access that has constrained previous implementations to thin samples. With an independent optimization at every scan position, our system extended localized optical focusing to scanning imaging through *ex vivo* tissue, marking a step toward translating PA-guided wavefront shaping into practical deep-tissue imaging configurations.

## FUNDING AND ACKNOWLEDGMENTS

This work was supported by the Engineering and Physical Sciences Research Council (EPSRC; EP/W524475/1), EPSRC 2023 Doctoral Training Partnership Studentship, EPSRC UKRI Innovation Fellowship (EP/S001506/1), and the Royal Academy of Engineering (RF2223-22-170). The funders had no role in study design, data collection and analysis, decision to publish, or preparation of the manuscript.

## AUTHOR CONTRIBUTIONS

X.L. and T.Z. conceived the research; X.L. built the experimental setup and conducted the experiments; X.L. performed the data analysis; X.L. wrote the manuscript. All authors contributed to the manuscript and approved the final version.

## DECLARATION OF INTERESTS

W.X., T.Z., and S.O. are inventors on a granted patent held by King's College London covering the real-valued intensity transmission matrix (RVITM) wavefront shaping algorithm used in this work (US 12,191,911 B2). No separate patent has been filed for the reflection-mode system reported here. The remaining authors declare no competing interests.

## DATA AND CODE AVAILABILITY

Data are available from the corresponding author upon reasonable request.

## ETHICAL STATEMENT

Not applicable.